# Intertwined nematic and $d$-wave superconductive orders in optimally-doped La$_{1.84}$Sr$_{0.16}$CuO$_4$

Gangfan Chen[1,2,3,4], Yichi Zhang[1,2,3,4], Guangyu Xi[1,2,3,4], JingYi. Shen[1,2,3,4], and Jie Wu[1,2,3,*]

[1]*Department of Physics, School of Science, Westlake University, Hangzhou 310024, China*

[2]*Research Center for Industries of the Future, Westlake University, Hangzhou31 0024, China*

[3]*Key Laboratory for Quantum Materials of Zhejiang Province, School of Science, Westlake University, Hangzhou, 310024, China*

[4]*School of Physics, Zhejiang University, Hangzhou 310027, China*

*Author to whom correspondence should be addressed: wujie@westlake.edu.cn

**Abstract:**

**The anisotropy of the superconducting state and superconducting fluctuations in the CuO$_2$ plane is directly related to the superconducting mechanism of copper oxide superconductors and is therefore pivotal for understanding high-temperature superconductivity. Here, we integrated the high-precision angle-resolved resistivity (ARR) measurement with a rotatable in-plane magnetic field to systematically study the angular dependence of superconducting fluctuations in optimally doped La$_{1.84}$Sr$_{0.16}$CuO$_4$ (LSCO). By independently controlling the directions of the current and the magnetic field, we are able to isolate the magneto-resistivity contributed by the superconducting vortex motion and distinguish excitations from nematic superconductivity and $d$-wave superconductive order based on their respective C$_2$ and C$_4$ symmetries. Signatures of two intertwined superconductive orders are also evident in the** **measured angular dependence of the critical current. A *T-B* phase diagram of different types of superconducting fluctuations is determined. These findings are closely related to other intriguing phenomena, such as pair density wave and charge density wave.**

The lattice structure of copper oxide superconductors consists of single or multiple $CuO_2$ planes sandwiched between charge reservoir layers, with high-temperature superconductivity residing at the $CuO_2$ planes [1-3]. $CuO_2$ planes in different layers are coupled via Josephson coupling, and the out-of-plane and in-plane directions exhibit significant anisotropy, making the superconductivity quasi-two-dimensional. Thus, the symmetry of the superconducting and normal states in the $CuO_2$ plane is crucial for understanding the superconducting mechanism. Unlike the normal state of conventional superconductors, which behaves like a Fermi liquid and is spatially isotropic, the normal state of copper oxide superconductors has been shown to spontaneously break the rotational symmetry and is anisotropic in the $CuO_2$ plane with **$C_2$** symmetry [4-10]. The nature of this electronic nematic normal state and its role in unconventional superconductivity remain fundamental open questions in the field.

Signatures of uniaxial orders, such as charge/spin stripe order [11-19] and pair density wave (PDW) [20-23], have been detected in copper oxide superconductors by various techniques, including neutron scattering [11, 12], scanning tunneling microscopy [18, 19, 21-23], optical pump-probe technique [15, 16], muon spin rotation spectroscopy [17], etc. In principle, intertwinement of the PDW and nematic order with the *d*-wave superconductive order [24, 25] allows the superconducting order parameter to take on various forms [26, 27], e.g., a nematic PDW state was identified in $Bi_2Sr_2CaCu_2O_{8+x}$ by Josephson tunneling microscopy [23]. To determine the actual order parameter and assess its universality across different families of copper oxide superconductors, it is imperative to elucidate the anisotropy of superconductivity within the $CuO_2$ plane in real space.

Magneto-transport has been utilized to characterize nematic superconductivity in Fe-based [28, 29] and nickelate superconductors [30]. The angular dependence of magneto-resistivity on the direction of the in-plane magnetic field reveals the anisotropy in the critical magnetic field and the spontaneous breaking of rotational symmetry in the superconducting state. These electrical transport experiments were carried out using the Carbino-shape configuration [29-31], so the measured magneto-resistivity was averaged over all the in-plane current directions. While this method is effective in retrieving the anisotropy induced by an external in-plane magnetic field, it overlooks the intrinsic anisotropy in resistivity corresponding to different current directions, e.g. correlated angular oscillations in longitudinal and transverse resistivity with continuously varying current direction due to electronic nematicity [10, 32].

To overcome this difficulty, we conceived and implemented the ARR method [32], which allows independent control of the in-plane directions of both excitation current and magnetic field. This enables us to distinguish the contributions to magneto-resistivity from various origins, such as electronic nematicity, the excitation of superconducting vortices, and superconducting fluctuations. The anisotropies of each effect were separated, and their correlations were mapped onto a *T-B* phase diagram. The observed **C$_2$** and **C$_4$** symmetry in superconducting fluctuation and critical current density suggest the coexistence of both nematic and *d*-wave superconductive order parameters.

Optimally doped 26 nm LSCO ($x = 0.16$) films were synthesized on LaSrAlO$_4$(001) (LSAO) substrates using the pulsed laser deposition technique. The high crystalline quality was confirmed by a clear *in situ* reflection high energy electron diffraction (RHEED) pattern, oscillations in RHEED spot intensity during deposition, and *ex situ* X-ray diffraction spectrum. A post-growth annealing of LSCO film was performed *in situ* at 450℃ for 40 minutes to remove excess oxygen. The LSAO surface before deposition and the LSCO film surface after deposition were atomic flat, verified by atomic force microscopy. The films were patterned by standard UV-lithography to create the structure shown in Fig. 1a for electric transport measurements. Further details on film synthesis, characterization, and fabrication are provided in the Methods section and the *Supplemental Material*.

To control the direction of current, we take advantage of the vector property of the current density ***J***. By independently controlling the *x* and *y* components of ***J***, $J_x$ and $J_y$, we can continuously rotate ***J*** in the *x-y* plane at the cross section of the pattern (Fig. 1a) (here the *x*- and *y*-axes correspond to the LSCO [100] and [010] directions, respectively) while simultaneously measure the corresponding voltages, $V_x$ and $V_y$.

Then the longitudinal and transverse resistivities with respect to **J**, $\rho$ and $\rho_T$, are then obtained from the corresponding $V_L$ and $V_T$ using the relations $V_L = V_x \sin\phi + V_y \cos\phi$ and $V_T = V_x \cos\phi - V_y \sin\phi$, where $\phi$ is the angle between **J** and the *x*-axis. Varying $\phi$ from 0° to 360° by rotating **J** in-plane yields the complete angular dependence of $\rho(\phi)$ and $\rho_T(\phi)$. The angular resolution is only limited by the output electronic precision so it can be unprecedented high (< 0.01°) if needed. Our sample holder is designed to be rotatable relative to the applied magnetic field, **B**, so that the direction of **B** can be varied continuously in-plane. In this way, the directions of **J** and **B** are controlled independently, enabling the anisotropic dependence of magneto-resistivity on each to be separated.

The effectiveness of the ARR methodology was verified in our previous publication [32]. There, a 10 nm thick gold control sample manifested faint angular oscillations ($\Delta R/\bar{R} < 0.05\%$) in the measured longitudinal and transverse resistance, $R(\phi)$ and $R_T(\phi)$, which were attributed to inevitable small film thickness variations. This demonstrated that the ARR method is highly sensitive in detecting anisotropy in electric transport. Additionally, similar lithography pattern and measurement scheme have been used to study directional superconducting vortex motion [33], or intentionally introduced anisotropy [34].

A sharp superconducting transition is observed in the synthesized LSCO (x = 0.16) film, with $T_c^{on}$ at 41.1 K, $T_c^{mid}$ at 40.6 K, and $T_c^{off}$ at 40.4 K (Fig. 1b). Here, $T_c^{on}$, $T_c^{mid}$, and $T_c^{off}$ are defined as the onset, midpoint, and offset temperatures corresponding to 90%, 50%, and 10% of the normal state resistivity, respectively. The narrow transition width and high $T_c$ value indicate the high quality and homogeneity of the film.

The angle-dependent $\rho(\phi)$ at zero magnetic field shows clear angular oscillations as **J** varies its direction in the CuO$_2$ plane (Fig. 1c). The 180° periodicity in $\phi$ evidences that the rotational symmetry is spontaneously broken and reduced to **C$_2$** symmetry. The $\rho(\phi)$ and correlated $\rho_T(\phi)$ oscillations originate from electronic nematicity and can be well fitted by the expressions $\rho(\phi) = \bar{\rho} + \Delta\rho\cos[2(\phi - \alpha)]$ and $\rho_T(\phi) = \Delta\rho\sin[2(\phi - \alpha)]$ (see the *Supplemental Material* for more details). Here $\bar{\rho} = (\rho_a + \rho_b)/2$ and $\Delta\rho = (\rho_a - \rho_b)/2$. $\rho_a$ and $\rho_b$ are the resistivities along two orthogonal symmetric axes associated with electronic nematicity, also referred to as the "nematic director".

Mathematically, $\rho_a$ and $\rho_b$ are the diagonal elements of the rank-2 resistivity tensor. $\alpha$ is the angle between the nematic director and the *x*-axis, as the nematicity is not necessarily aligned with the lattice. The deviation of these expressions and studies on electronic nematicity in the normal state of LSCO can be found in our previous publications [10, 32] and the *Supplemental Material*. The focus of the current work, however, is on the transition from the superconducting to the normal state and the vestigial order in superconducting fluctuations.

The peaks in $\rho(\phi)$ shift abruptly as *T* lowers across $T_c^{on}$ (indicated by the dashed line in Fig. 1c). Concomitantly, the amplitude of $\rho(\phi)$ oscillations is augmented significantly for $T < T_c^{on}$. Both facts imply that the underlying mechanisms responsible for electronic nematicity are different for the normal and superconducting fluctuating (SF) states. The enhanced nematicity in the SF state is attributable to the vestigial nematicity in superconductivity.

The nematic nature of superconductivity is collaborated by the *I-V* curves taken below $T_c^{off}$. From the representative *I-V* curves for 30º ≤ $\phi$ ≤ 100º, it is evident that the critical current $I_c$ varies appreciably with $\phi$ (Fig. 1d). The retrieved $I_c(\phi)$ is strongly anisotropic, with the maximum $I_c$ being roughly twice the minimum $I_c$ (Fig. 1e). $I_c(\phi)$ can be well fitted by the expression $I_c(\phi) = a/(b + cos(2\phi - \alpha))$, where a, b and $\alpha$ are the fitting parameters. Moreover, $I_c(\phi)$ and $\rho(\phi)$ are closely correlated: both have a 180º periodicity in $\phi$, and the peaks in $I_c(\phi)$ ($\phi$ = 110º and 290º) correspond to the valleys in $\rho(\phi)$. The lower resistivity and higher critical current imply that superconductivity is most robust along $\phi$ = 110º.

Applying a magnetic field suppresses superconductivity and exposes the hidden states below $T_c^{off}$. In the presence of the magnetic field, the width of superconducting transition broadens and $T_c^{off}$ is lowered to 37.6 K at 16 T field (Fig. 2a), the maximum magnetic field accessible in our lab. The motion of superconducting vortex dissipates energy and is resistive to current flow. To distinguish this contribution from others, e.g., quasiparticle excitation and superconducting phase fluctuation, we measured the angle-dependent magneto-resistivity $\rho(\phi, \varphi)$ using three different schemes. Here, $\varphi$ is the angle between ***B*** and *x* axis, while ***B*** is always kept in-plane.

In the first scheme (Fig. 2b), $\boldsymbol{J}$ is fixed to be along $x$ axis and $\varphi$ changes from 0° to 360°. The Lorentz force acted on superconducting vortices by the magnetic field is $F_L = \boldsymbol{J} \times \boldsymbol{B}/c$, so the induced vortex motion $\boldsymbol{v}$ is normal to the film [35]. Consequently, this induces an electric field of magnitude $\boldsymbol{E} = \boldsymbol{B} \times \boldsymbol{v}/c$ that is in-plane and transverse to $\boldsymbol{B}$. Thus, the resultant longitudinal magneto-resistivity along $\boldsymbol{J}$ should have the relation $\rho(\phi = 0°, \varphi) \propto sin^2\varphi$. This is consistent with the measured $\rho(\phi = 0°, \varphi)$ at variable $B$ (Fig. 2b). $\rho(\phi = 0°, \varphi)$ is dominated by $sin^2\varphi$ oscillation, whose amplitude declines with $B$. In comparison, $\rho(\phi = 0°, \varphi)$ of the normal state, which is free of superconducting vortex, shows negligible angular oscillations (Fig. 2e). Thus, this confirms that the $\rho(\phi = 0°, \varphi)$ oscillations originate from vortex motion. More measurements and discussions on $\rho(\phi = 0°, \varphi)$ are provided in the Supplemental Material.

This scheme is widely used in in-plane angular magnetoresistance (AMR) measurement, where the current flow is fixed in direction in a Hall-bar geometry and the magnetic field is rotated in-plane relative to the current. A twofold in-plane AMR was reported in underdoped LSCO [36], YBa$_2$Cu$_3$O$_{6+x}$ [13], and electron-doped La$_{2-x}$Ce$_x$CuO$_4$ [37] while a fourfold AMR was found in lightly electron-doped Pr$_{1.3-x}$La$_{0.7}$Ce$_x$CuO$_4$ [38], Nd$_{2-x}$Ce$_x$CuO$_4$ [39], and Pr$_{2-x}$Ce$_x$CuO$_4$ [40]. The twofold AMR is suggested to be associated with spin ordering, whereas the fourfold AMR is attributed to a magnetic-field-induced spin flop transition. These findings, however, are fundamentally different from the $sin^2\varphi$ oscillations shown in Fig. 2b. The AMR results mentioned above pertain to underdoped cuprates, most of which are non-superconducting. In contrast, the LSCO film in our study is optimally doped. More importantly, the twofold or fourfold AMR observed in previous studies occurs at temperatures much higher than $T_c$ and diminishes with the static or quasistatic antiferromagnetic order [40]. Conversely, $sin^2\varphi$ oscillations in our LSCO film are substantial in amplitude only near $T_c$ and disappear completely once superconducting fluctuation vanishes (Fig. 2e). Therefore, the $sin^2\varphi$ oscillations should be associated with superconducting vortices rather than antiferromagnetic order.

In order to probe the symmetry of superconducting order parameters, it is necessary to fix the contribution of mobile superconducting vortex to magneto-resistivity by fixing the relative angles between $\boldsymbol{J}$ and $\boldsymbol{B}$. Hence, the second measurement scheme is to keep $\boldsymbol{B}$ perpendicular to $\boldsymbol{J}$ and rotate both of them simultaneously so that the Lorentz force due to the magnetic field remains maximized (Fig. 2c). The third scheme, on the other hand, is to keep **B** parallel to **J** during rotation to completely remove the influences of the Lorentz force (Fig. 2d). The measured $\rho(\phi, \varphi = \phi + 90°)$ and $\rho(\phi, \varphi = \phi)$ show remarkable differences at low magnetic fields but manifest similar $sin\,(2\phi + 114°)$ oscillations at high magnetic fields (Figs. 2c and 2d). It is worth noting that the peak angles ($\phi = 168°$ and $348°$) in $\rho(\phi, \varphi = \phi + 90°)$ differ from those in $\rho(\phi = 0°, \varphi)$ but identical to those in $\rho(\phi, \varphi = \phi)$ at high magnetic fields, e.g., $B$ = 14 T. Moreover, they are also the same as the peak angles in $\rho(\phi)$ at 41 K in the absence of a magnetic field. This is no coincidence. It illustrates that $\rho(\phi, \varphi = \phi + 90°)$ and $\rho(\phi, \varphi = \phi)$ at large **B**-fields are dominated by their dependences on **J** and, therefore, are governed by electronic nematicity in the normal state. Once superconducting fluctuations are suppressed by elevated temperatures, or magnetic fields, or both, $\rho(\phi, \varphi = \phi + 90°)$ and $\rho(\phi, \varphi = \phi)$ are dictated by the nematicity in the normal state, which is insusceptible to the in-plane magnetic field (Fig. 2f).

As the magnetic field decreases and superconducting fluctuations strengthen, two dips at $\phi$ = 102° and 260° gradually develop in $\rho(\phi, \varphi = \phi + \pi/2)$ (Fig. 2c). This seems to contradict the expectation that the magneto-resistivity induced by vortex motion, being proportional to $\bm{J} \times \bm{B}$, should be angle-independent in the second measurement scheme. It may be related to the drifting of vortices along the current flow, which could be anisotropic in-plane and influenced by randomly distributed pinning centers along the current's path. Further investigations are required to elucidate the mechanism underlying this observation.

In the absence of the Lorentz force and the resultant vortex motion, the measured $\rho(\phi, \varphi = \phi)$ reflects the symmetry of superconducting order parameter. As $B$ decreases, the symmetry of $\rho(\phi, \varphi = \phi)$ changes from twofold to fourfold (Fig. 2d), e.g. at $B$ = 3 T, the period of angular oscillation becomes 90°. Applying a stronger magnetic field shifts $T_c^{off}$ to lower temperature. Following the superconducting-to-normal phase boundary $T_c^{off}(B)$, the four-fold oscillations are found to be ubiquitous in the vicinity of superconducting transition. For instance, the $\rho(\phi, \varphi = \phi)$ taken at $T$ = 38.2 K and $B$ = 14 T (Fig. 3a) shows four peaks at $\phi$ = 45°, 135°, 225°, and 315° that correspond to the Cu-O-Cu bond directions of LSCO lattice.

In accord with $\rho(\phi, \varphi = \phi)$ in the SF state, the critical current $I_c(\phi, \varphi = \phi)$ for the superconducting state, with ***B*** parallel to ***J***, manifests a mixture of twofold and fourfold symmetries (Fig. 3b), in contrast to the twofold symmetric $I_c(\phi)$ in the absence of magnetic field (Fig. 1e). The dips of $I_c(\phi, \varphi = \phi)$ occur at $\phi = 45°, 135°, 225°$, and $315°$ (Fig. 3b) that have one-to-one correspondence with the peaks in $\rho(\phi, \varphi = \phi)$ (Fig. 3a). Thus, superconductivity is more robust along the Cu-Cu bond than the Cu-O-Cu bond direction, consistent with the widely accepted $d_{x^2-y^2}$ superconducting gap symmetry for copper oxide superconductors.

In addition to the **C₄** component, it is evident that $I_c(\phi, \varphi = \phi)$ also contains a **C₂** component, e.g. two well-defined peaks in $I_c(\phi, \varphi = \phi)$ emerge at $\phi = 110°$, and $290°$ (Fig. 3b), which are the same angles corresponding to the valleys in $\rho(\phi)$ at $T = 40.1$ K (Fig. 1c). Hence, the **C₂** component originates from the electronic nematicity in the SF state. The coexistence of **C₄** and **C₂** components implies that nematic superconductivity coexists with *d*-wave superconductivity in LSCO.

To map out an explicit phase diagram for states with **C₂** and **C₄** symmetries, we retrieved the amplitudes $\Delta\rho_{2\phi}$ and $\Delta\rho_{4\phi}$ from the best fits of the resistivity by the expression $\rho(\phi, \varphi = \phi) = \bar{\rho} + \Delta\rho_{2\phi}\cos[2(\phi - \alpha)] + \Delta\rho_{4\phi}\cos[4(\phi - \beta)]$ (Figs. 3c and 3d), where $\alpha$ and $\beta$ are two fitting parameters standing for the phase offsets. In the *T-B* phase diagram, $T_c^{on}(B)$ and $T_c^{off}(B)$ denote the phase boundaries separating the superconducting, SF, and normal states. Remarkably, the phases with appreciable $\Delta\rho_{4\phi}$ closely trace the $T_c^{on}(B)$ curve, implying that the **C₄** component emerges only in the vicinity of the superconducting state. The **C₂** component $\Delta\rho_{2\phi}$, on the other hand, is weak where the **C₄** component is present but persists all the way to high *T* and *B*.

The amplitude of fourfold angular oscillation, $\Delta\rho_{4\phi}$, is a measure of the amplitude of *d*-wave order parameter. In magneto-transport experiment, the fourfold symmetry was observed in several superconducting copper oxides and was ascribed to the *d*-wave superconducting gap symmetry. Examples include the *c*-axis magneto-resistivity with variable in-plane magnetic field in $Pb_2Sr_2Y_{0.62}Ca_{0.38}Cu_3O_8$ [41] and $La_{1.6-x}Nd_{0.4}SrCuO_4$ [42], in-plane torque anisotropy in $Tl_2Ba_2CuO_{6+\delta}$ [43], and angular dependence of the upper critical field in LSCO [44]. In other superconductors, the fourfold symmetry manifested in magnetoresistance was also attributed to their superconducting symmetry, e.g., in $LiTi_2O_4$ [31] and $Nd_{0.8}Sr_{0.2}NiO_2$ [30]. Thus, it is no surprise that the fourfold symmetry emerges in $\rho(\phi, \varphi = \phi)$ of our optimally doped LSCO and $\Delta\rho_{4\phi}$ serves as an indicator for the strength of the *d*-wave superconductive order.

The amplitude of twofold angular oscillation, $\Delta\rho_{2\phi}$, is a measure of the order parameter of nematic superconductivity. Although a direct measurement of the nematicity in the superconducting state is not feasible via electric transport, superconducting fluctuation at the temperatures between $T_c^{on}(B)$ and $T_c^{off}(B)$ is clearly nematic that is an imminent consequence of nematic superconductivity. The nematicity may be related to the stripe phase or checkerboard order [11-19] and can couple to pair density wave order to form nematic pair density wave, as indicated by the recent Josephson tunneling microscopy [23].

The phase diagram of the superconductive orders is summarized in Fig. 4a. As *T* or *B* increases, the superconducting state evolves into the intermediate state with *d*-wave superconductive order fluctuations ("DSC fluctuating state"), then the intermediate state with nematic superconducting fluctuations ("NSC fluctuating state"), and eventually the nematic strange metal state ("NSM state"). The polar plots of $\rho(\phi, \varphi = \phi)$ for these three states differ drastically. The "cloverleaf" shape of $\rho(\phi, \varphi = \phi)$ for the DSC fluctuating state is aligned with the crystalline lattice, with $\rho(\phi, \varphi = \phi)$ peaking along the Cu-O-Cu bond direction (Figs. 4f and 4g). Conversely, the directors of the "dumbbell" shape of $\rho(\phi, \varphi = \phi)$ for the NSC fluctuating state (Figs. 4d and 4e) and the NSM state (Figs. 4b and 4c) are not aligned with the lattice. Notably, there is an abrupt rotation of the nematic director across the boundary between the NSC fluctuating state and the NSM state.

From these results, we conclude that nematic and *d*-wave superconductive orders coexist in high temperature superconducting copper oxides. The intertwining of electronic nematicity and superconductivity enhances the amplitude of nematicity and alters its director, as evidenced by the stronger nematicity in the NSC state compared to the NSM state. This finding provides new insights into the pairing mechanism and the formation of phase coherence in high-temperature superconductivity.

**Acknowledgements**

This work was supported by National Key R&D Program of China (2023YFA1406400 and No. 2024YFA1408102), Research Center for Industries of the Future (RCIF project No. WU2023C001) at Westlake University, the National Natural Science Foundation of China (Grant No. 12174318), the Zhejiang Provincial Natural Science Foundation of China (Grant No. XHD23A2002). We acknowledge the assistance provided by Dr. Chao Zhang of the Instrumentation and Service Center for Physical Sciences and the Westlake Center for Micro/ Nano Fabrication at Westlake University.

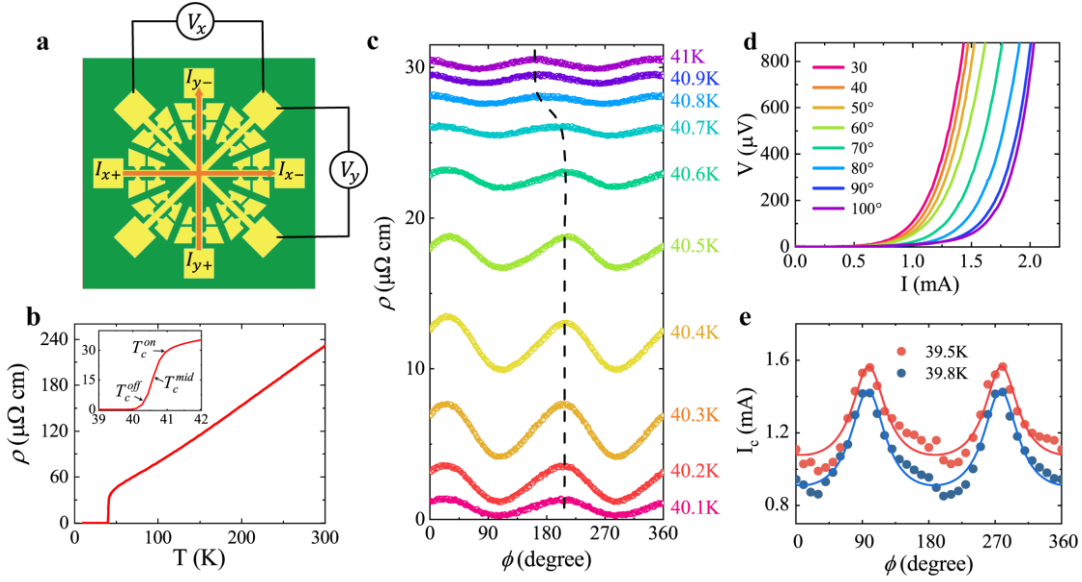

**Figure 1 | Anisotropic electric transport of LSCO in the absence of magnetic field.
a,** The schematic drawing of the pattern used for the ARR measurements. The $x$ and $y$ components of the current density $\boldsymbol{J}$, $J_x$ and $J_y$, are independently controlled by the currents injected into the paired contacts $I_{x+}$ and $I_{x-}$, and the pair $I_{y+}$ and $I_{y-}$, respectively. Thus, $\boldsymbol{J}$ at the cross section of the pattern can be rotated continuously in-plane while the $x$- and $y$-components of the voltages, $V_x$ and $V_y$, are measured simultaneously. **b,** The temperature-dependent longitudinal resistivity $\rho(T)$, measured along the LSCO [100] direction, manifests sharp superconducting transition. The inset magnifies the superconducting transition with $T_c^{on}$ at 41.1 K, $T_c^{mid}$ at 40.6 K, and $T_c^{off}$ at 40.4 K. Here $T_c^{on}$, $T_c^{mid}$, and $T_c^{off}$ denote the temperatures with 90%, 50%, and 10% of the normal state resistivity. **c,** The angle-dependent $\rho(\phi)$ at variable $T$ around the superconducting transition evidence the electronic nematicity. Here $\phi$ is the angle between $\boldsymbol{J}$ and the LSCO crystalline [100] direction. The dashed line traces the peaks in $\rho(\phi)$, illustrating the transition of nematic director due to nematic superconducting fluctuations. **d,** The $I$-$V$ curves at a fixed temperature ($T$ = 39.5 K) taken at systematically varying $\phi$ clearly show that the critical current $I_c$ depends on the angle $\phi$. **e,** The angle-dependent $I_c(\phi)$ at two representative temperatures manifest that the superconductivity is strongly anisotropic in-plane and its **C₂** symmetry is correlated with the nematicity, i.e., the peaks in $I_c(\phi)$ corresponds to the valleys in $\rho(\phi)$ at $T$ = 40.1 – 40.7 K range. The solid circles are experimental data and the solid curves are the best fittings according to the expression $I_c(\phi) = a/(b + cos(2\phi - \alpha))$.

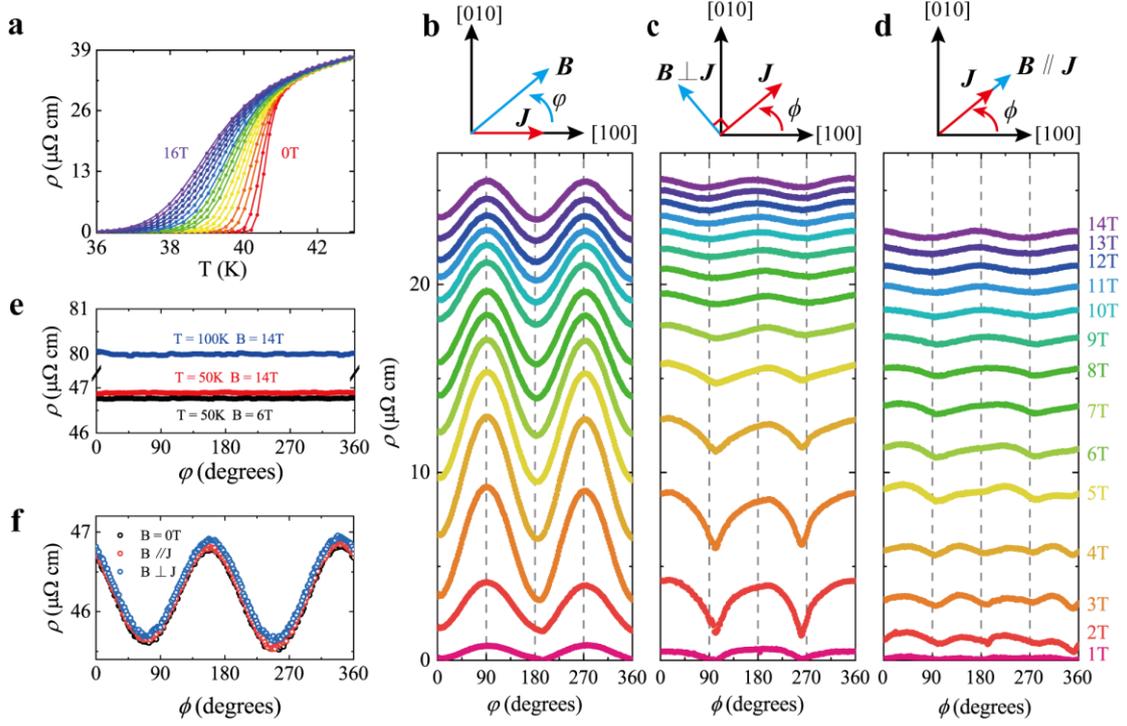

**Figure 2 | Anisotropic magneto-resistivity of LSCO. a,** The resistivity $\rho(T)$ under incremental magnetic field along *x*-axis (LSCO [100] direction) from 0 to 16 T in 1 T step. **b,** The longitudinal resistivity $\rho(\phi = 0°, \varphi)$ under varying magnetic fields ***B*** from 1 to 14 T in 1 T step. The angle $\varphi$ between ***B*** and *x*-axis varies in-plane while the current flow ***J*** is fixed to be along the *x*-axis. The dominating $sin^2\varphi$ oscillations originate from vortex motions, driven by the Lorentz force on superconducting vortices. **c,** With ***J*** perpendicular to ***B*** (***J***⊥***B***), i.e., $\varphi = \phi + 90°$, $\rho(\phi, \varphi = \phi + 90°)$ are under the maximized influence of the Lorentz force. **d,** With ***J*** in parallel to ***B*** (***J***//***B***), i.e., $\varphi = \phi$, the Lorentz force diminishes. $\rho(\phi, \varphi = \phi)$ manifest fourfold symmetry at low magnetic field (1 T ≤ *B* ≤ 4 T), and twofold symmetry at higher magnetic field (5 T ≤ *B* ≤ 14 T). The temperature for panel **b**, **c**, and **d** is 40 K that is below $T_c^{off}$ at zero magnetic field. **e,** As a comparison, $\rho(\phi = 0°, \varphi)$ shows no angular dependence on $\varphi$ for the normal state of LSCO. **f,** $\rho(\phi, B = 0)$ at zero *B*-field has the same dependence on $\phi$ as $\rho(\phi, \varphi = \phi + 90°)$ and $\rho(\phi, \varphi = \phi)$ taken at *B* = 5 T for the ***J***⊥***B*** and ***J***//***B*** measurement schemes, respectively. The temperature, *T* = 50 K, is the same for all plots. This clearly shows that once the LSCO film is in the normal state, $\rho(\phi, B = 0)$, $\rho(\phi, \varphi = \phi + 90°)$ and $\rho(\phi, \varphi = \phi)$ are all dictated by the electronic nematicity of the normal state and is insensitive to the direction of ***B***.

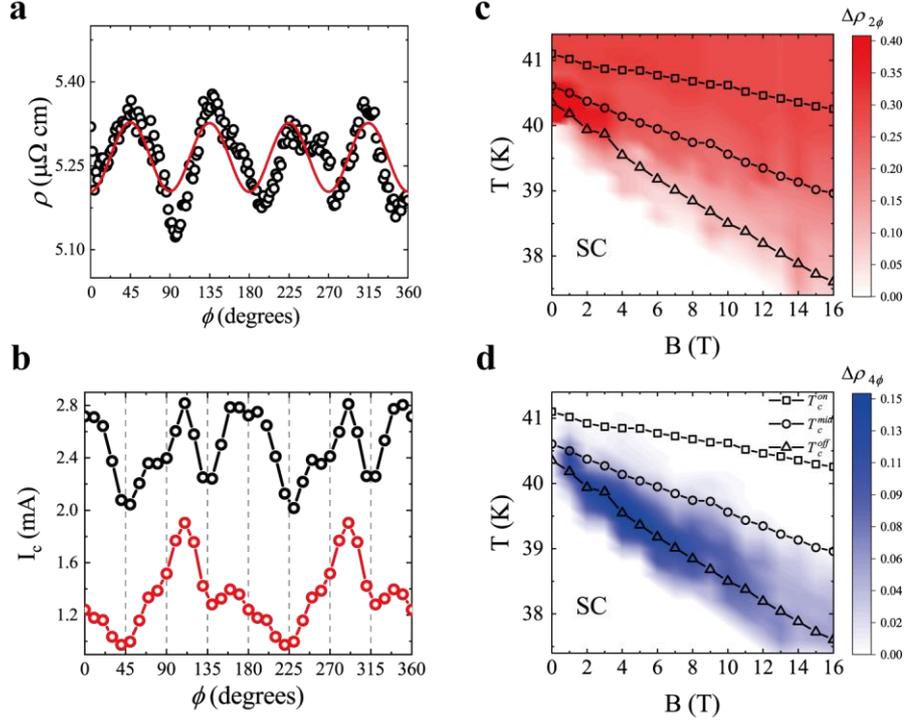

**Figure 3 | B-T phase diagram for the C₂ and C₄ anisotropies in superconducting fluctuations. a,** The angular oscillations in $\rho(\phi, \varphi = \phi)$ taken at $T = 38.2$ K and $B = 14$ T with ***J//B***. The experimental data (black open circles) can be well fitted by the expression $\rho(\phi, \varphi = \phi) = \bar{\rho} + \Delta\rho\cos[4(\phi - \alpha)]$ (red lines). **b,** The critical current $I_c(\phi, \varphi = \phi)$ taken at $T = 37$ K (black open circles) and 37.5 K (red open circles) at $B = 8$ T. **c**, and **d**, The *B-T* phase diagram for the **C₂** and **C₄** components of the angle-dependent resistivity $\rho(\phi, \varphi = \phi)$. The amplitudes $\Delta\rho_{2\phi}$ and $\Delta\rho_{4\phi}$ are retrieved from the expression $\rho(\phi, \varphi = \phi) = \bar{\rho} + \Delta\rho_{2\phi}\cos[2(\phi - \alpha)] + \Delta\rho_{4\phi}\cos[4(\phi - \beta)]$. The $T_c^{on}(B)$, $T_c^{mid}(B)$, and $T_c^{off}(B)$ curves denote the temperatures with 90%, 50%, and 10% of the normal state resistivity, respectively. "SC" stands for the superconducting state.

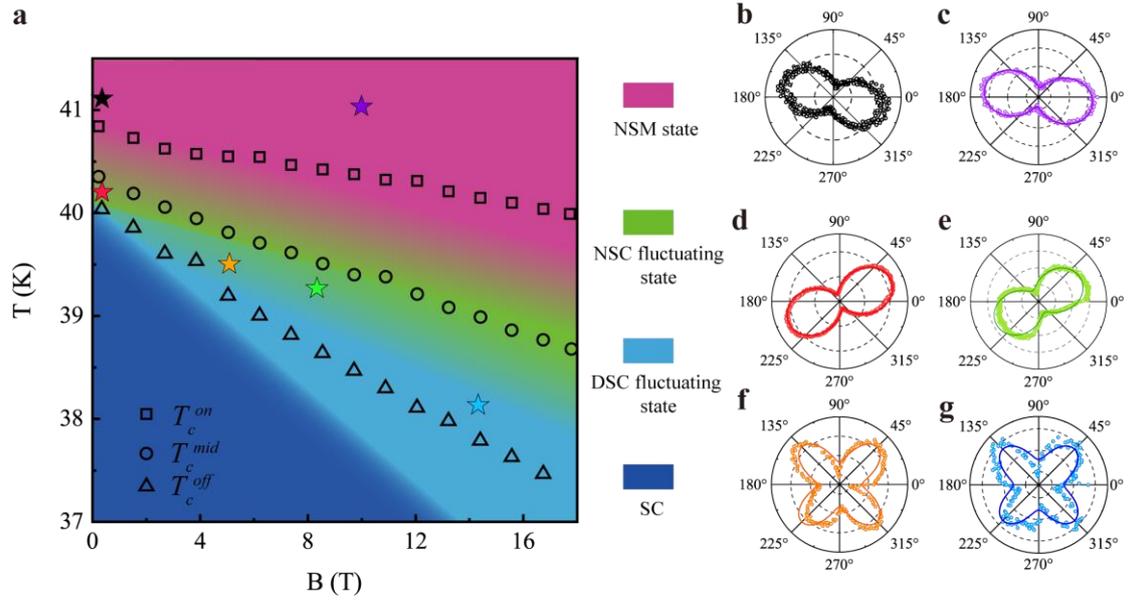

**Figure 4 │ *B-T* phase diagram for superconducting fluctuations. a**, "SC" stands for the superconducting state, "DSC fluctuating state" for the intermediate state with *d*-wave superconductive order fluctuations, "NSC fluctuating state" for the intermediate state with nematic superconducting fluctuations, and "NSM state" for the nematic strange metal state. **b**-**g**, The polar plots of $\rho(\phi, \varphi = \phi)$ show the directors of $C_2$ and $C_4$ anisotropies at representative (*T*, *B*) values, with their corresponding locations marked by stars in panel **a**. The colors of the stars correspond one-to-one with the colors of the respective polar plots.

# Intertwined nematic and *d*-wave superconductive orders in optimally-doped La$_{1.84}$Sr$_{0.16}$CuO$_4$

G.F. Chen[1,2,3,4], Y.C. Zhang[1,2,3,4], G.Y. Xi[1,2,3,4], J.Y. Shen[1,2,3,4], J. Wu[1,2,3,*]

[1]Department of Physics, School of Science, Westlake University, Hangzhou 310024, China

[2]Research Center for Industries of the Future, Westlake University, Hangzhou31 0024, China

[3]Key Laboratory for Quantum Materials of Zhejiang Province, School of Science, Westlake University, Hangzhou, 310024, China

[4]School of Physics, Zhejiang University, Hangzhou 310027, China

*Author to whom correspondence should be addressed: wujie@westlake.edu.cn

## Supplemental materials

**1. Film synthesis, device fabrication and electrical transport measurement**

The La$_{1.84}$Sr$_{0.16}$CuO$_4$(001) thin films were epitaxially grown on SrLaAlO$_4$(001) substrates by pulsed laser deposition using a 248 nm KrF laser. Prior to growth, the SrLaAlO$_4$ substrate was annealed at 720°C for 20 mins under an oxygen pressure of 10$^{-2}$ Torr. During deposition, the substrate temperature was kept at 700°C with the oxygen pressure of 9×10$^{-2}$ Torr. A laser fluence of 1 J/cm$^2$ was used to ablate the target at a repetition frequency of 2 Hz. For post-annealing, the LSCO films were held at 450°C for 40 mins in an oxygen pressure of 3.8×10$^2$ Torr.

The films were patterned by standard UV-lithography and ion milling to form the device shown in Fig. 1a. Then the electrodes (Au, 100 nm) were deposited via magneton sputtering. X-ray diffraction experiments were conducted at room temperature using a Bruker D8-Advanced X-ray diffractometer with Cu $K_\alpha$ radiation ($\lambda$ = 1.5406 Å).

The electrical transport measurements were performed using a 16 T Physical Property Measurement System (PPMS-16T, Quantum Design) equipped with a rotator featuring an angular accuracy of 0.05°. For transport measurements, DC currents were supplied by Keithley 2450 sourcemeters, and DC voltages were measured using Keithley 2182A nanovoltmeters.

## 2. Structural and morphological characterizations

Structural and morphological characterizations confirm the high quality of synthesized single-crystalline $La_{1.84}Sr_{0.16}CuO_4$ (LSCO) films.

The crystalline structure of the LSCO film is monitored *in situ* in real time using the reflection high energy electron diffraction (RHEED) apparatus during film deposition. The streak-like RHEED pattern and the oscillations in RHEED spot intensity over time (Fig. S1) confirm that the film grows in the desired layer-by-layer mode and that the LSCO film is single-crystalline with a flat surface throughout the growth process. It is remarkable that the RHEED spot intensity does not decay even after 20 cycles of oscillations—a fingerprint of ideal film synthesis.

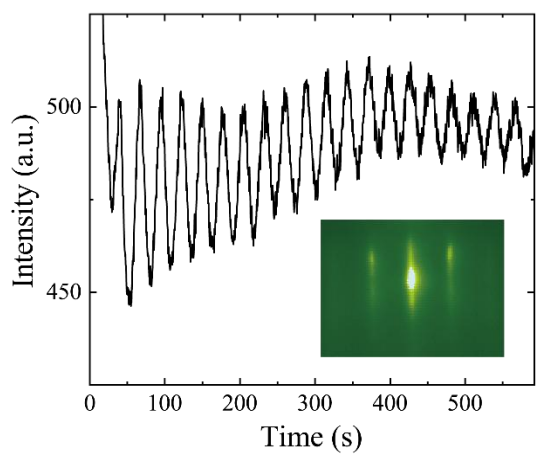

**Figure S1 | The oscillations in RHEED spot intensity over deposition time.** The intensity does not decay even after 20 cycles of oscillations. The inset shows the RHEED pattern during growth.

The lattice structure of the LSCO film is further examined *out situ* using X-ray diffraction (XRD). The pronounced XRD peak corresponding to LSCO (006) diffraction peak is manifested in the $2\theta$-$\omega$ scan (Fig. S2), confirming the single-crystallinity of the LSCO film. The measured *c*-axis lattice constant is 1.326 nm. The periodic fringe peaks reflect that both the LSCO/LaSrAlO4 interface and the LSCO surface are sharp enough to produce the interference of X-rays. The film thickness, determined from the periodicity of the fringe peaks, is 26 nm that agrees nicely with the thickness measured by RHEED oscillations.

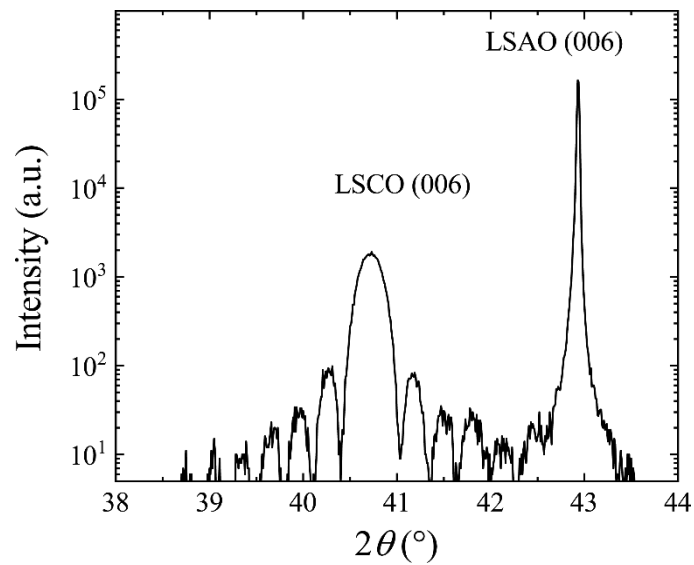

**Figure S2 │ The XRD $2\theta$-$\omega$ scan of the LSCO film.**

The morphology of the LSCO film is studied using atomic force microscopy (AFM). The AFM image of a 20 uc (~26 nm) thick LSCO film shows clear atomic steps (Fig. S3). The overall root-mean-square (r.m.s.) surface roughness is about 0.377 nm, which should be contrasted with the *c*-axis lattice constant, 1.326 nm. Thus, the surface roughness is much less than the height of a single unit cell, implying that the LSCO film is atomically flat.

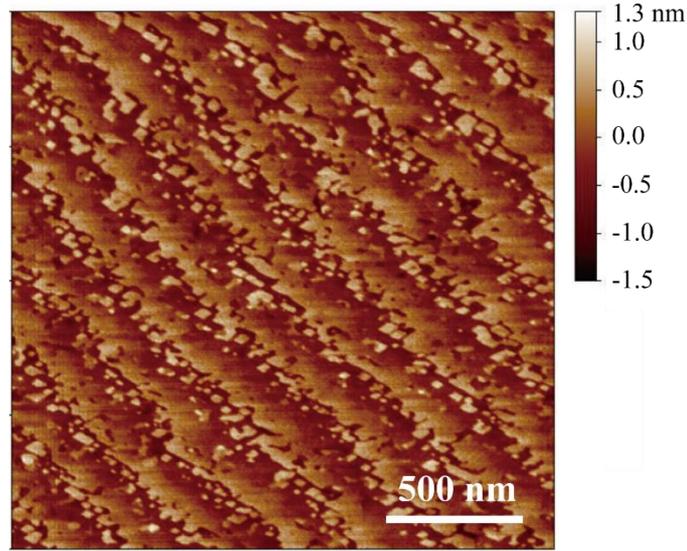

**Figure S3 │ AFM image taken from a 26 nm thick LSCO film.** Atomic steps are clearly visible. The rms roughness is about 0.377 nm.

## 3. Homogeneity of superconductivity

The sharpness of superconducting transition is a sensitive measure of the homogeneity of superconductivity. We facilitate the two-coil mutual inductance technique[45-47] to measure the diamagnetic signal due to Meissner effect.

The voltage across the pick-up coil, $V_p$, have both in-phase and off-phase components, $\mathrm{Re}V_p$ and $\mathrm{Im}V_p$, in response to the *ac* currents in the drive coil, which generates an *ac* magnetic field penetrating through the pick-up coil. $\mathrm{Re}V_p$ drops abruptly as soon as the LSCO film enters the superconducting state, and the screening supercurrent is excited by the *ac* magnetic field (Fig. S4). Simultaneously, $\mathrm{Im}V_p$ manifests a pronounced peak right below $T_c$, due to energy dissipation consumed by superconducting fluctuations. The full width at half maximum (FWHM) of the $\mathrm{Im}V_p(T)$ peak is around 1 K, which is a good estimate of $T_c$ spread in our 5 mm × 5mm LSCO film.

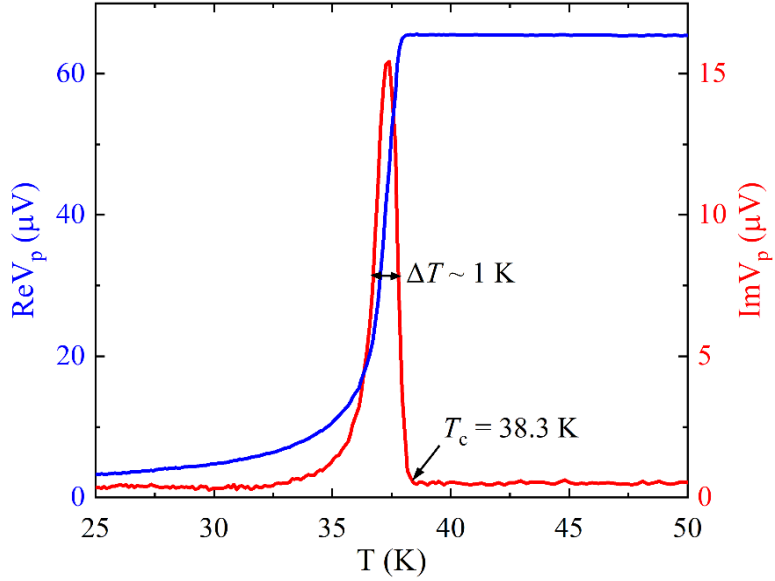

**Figure S4 | Homogeneity of superconductivity revealed by the mutual inductance measurement on a 26 nm thick LSCO film.** $V_p$ is the voltage across the pick-up coil, which is proportional to the mutual inductance. The real and imaginary components of $V_p$, $\text{Re}V_p$ and $\text{Im}V_p$, correspond to the blue and red curves line respectively. The FWHM of the $\text{Im}V_p(T)$ peak, $\Delta T$, is 1 K and the superconducting temperature, $T_c$, is 38.3 K.

## 4. Electronic nematicity of the normal state

The anisotropic electrical transport due to electronic nematicity in the normal state of LSCO was elucidated in our previous publications[10, 32]. This is not the main focus of the current work, but, for the completeness of the work, here we present a brief summary together with the measurement results on the LSCO film, from which the data in Figs. 1-3 were collected.

The in-plane resistivity is a rank-2 tensor:

$$\rho = \begin{pmatrix} \rho_a & 0 \\ 0 & \rho_b \end{pmatrix},$$

where $\rho_a$ and $\rho_b$ are resistivities along two principal axes $\mathbf{e}_a$ and $\mathbf{e}_b$, respectively. If $\rho_a = \rho_b$, then $\rho$ is simply a scalar multiplied by a unit matrix. In the presence of electronic nematicity, the in-plane rotational symmetry is spontaneously reduced to **C₂** symmetry. Consequently, $\rho_a \neq \rho_b$. $\mathbf{e}_a$ and $\mathbf{e}_b$ are orthogonal to each other, and they are the symmetric axes of **C₂** symmetry.

For currents applied in the direction making an angle $\phi$ from the $\mathbf{e}_a$ axis, the

resistivity tensor becomes

$$\rho' = \hat{C}_\phi \rho \hat{C}_\phi^{-1} = \begin{pmatrix} \cos\phi & -\sin\phi \\ \sin\phi & \cos\phi \end{pmatrix} \begin{pmatrix} \rho_a & 0 \\ 0 & \rho_b \end{pmatrix} \begin{pmatrix} \cos\phi & \sin\phi \\ -\sin\phi & \cos\phi \end{pmatrix}$$

$$= \begin{pmatrix} \bar{\rho} + \Delta\rho\cos(2\phi) & \Delta\rho\sin(2\phi) \\ \Delta\rho\sin(2\phi) & \bar{\rho} - \Delta\rho\cos(2\phi) \end{pmatrix}.$$

Here $\bar{\rho} = (\rho_a + \rho_b)/2$ and $\Delta\rho = (\rho_a - \rho_b)/2$.

Then the longitudinal and transverse resistivity are given by

$$\rho(\phi) = \bar{\rho} + \Delta\rho\cos(2\phi),$$

$$\rho_T(\phi) = \Delta\rho\sin(2\phi).$$

Taking into consideration that the principal axes $e_a$ and $e_b$, in general, are not aligned with the crystallographic orientations [100] and [010], we denote the angle between $e_a$ and [100] as $\alpha$. Then the expressions of $\rho(\phi)$ and $\rho_T(\phi)$ become:

$$\rho(\phi) = \bar{\rho} + \Delta\rho\cos[2(\phi - \alpha)] \quad \ldots (1)$$

$$\rho_T(\phi) = \Delta\rho\sin[2(\phi - \alpha)] \quad \ldots (2)$$

These expressions[10, 32] agree nicely with experimental results.

For the LSCO film, from which the data in Figs. 1-3 were collected, $\rho(\phi)$ and $\rho_T(\phi)$ were measured for its normal state. The angular oscillations in $\rho(\phi)$ and $\rho_T(\phi)$ at $T$ = 300 K can be simultaneously fitted by Expressions (1) and (2) using the same set of fitting parameters $\Delta\rho$ and $\alpha$ (Fig. S5a). The nematicity magnitude can be represented by $N \equiv \Delta\rho/\bar{\rho}$. And $N(T)$ increases as $T$ decreases (Fig. S5b), consistent with the expectation that the nematicity will strengthen as the thermal fluctuations get weaker at lower temperatures. The electronic nematicity behaves very differently in the presence of strong superconducting fluctuations as $T$ approaches $T_c$. $\rho_T(\phi)$, poltted in the polar coordinate system, intuitively show the "cloverleaf" shape for both $T$ = 300 K and $T$ = $T_c$ (Fig. S5c). Moreover, there is a relative rotation between two cloverleaf patterns, and the angular difference, $\Delta\alpha = |\alpha(T = 300K) - \alpha(T = T_c)|$, is about 60°. This is consistent with our previous findings[10, 32], implying that superconducting fluctuations are strongly nematic—a direct evidence for nematic superconductivity in LSCO.

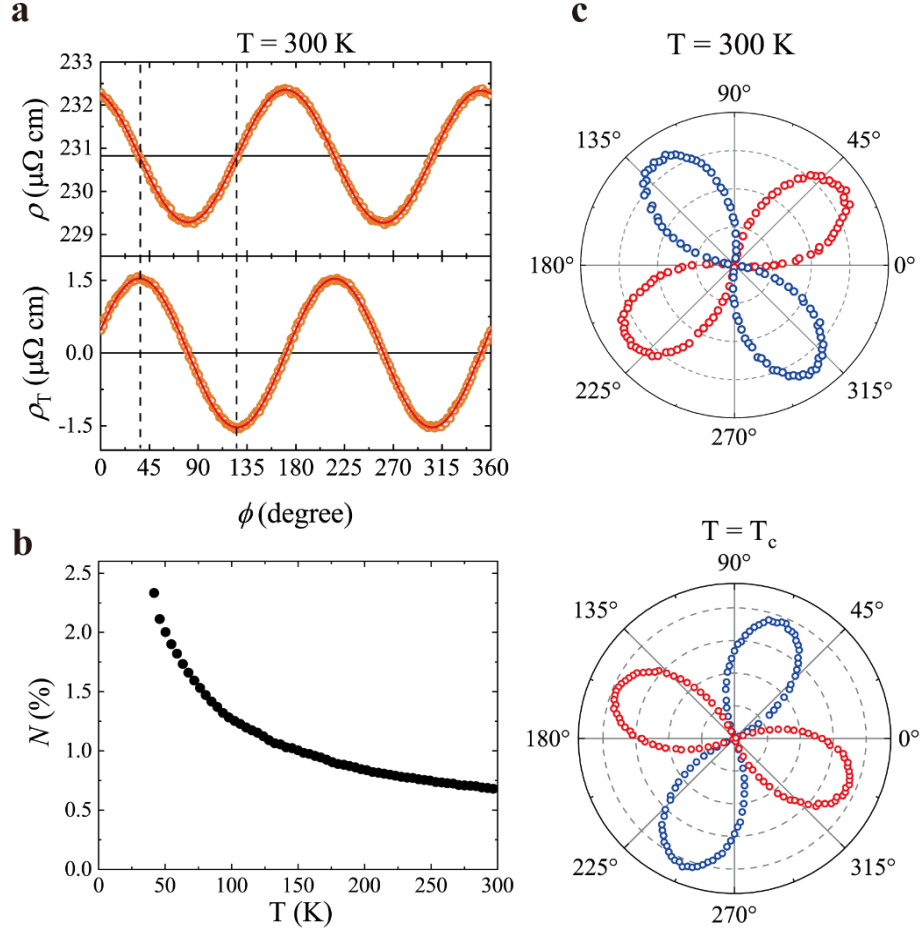

**Figure S5 | Electronic nematicity in LSCO film by ARR measurements. a**, $\rho(\phi)$ and $\rho_T(\phi)$ manifest angular oscillations with 180° period—a reflection of the $C_2$ symmetry of electronic nematicity. The open dots are experimental data and the red lines are the best fittings using Expressions (1) and (2). **b**, The temperature dependence of nematicity magnitude, $N \equiv \Delta\rho/\bar{\rho}$. **c**, $\rho_T(\phi)$ plotted in the polar coordinate system at $T = 300$ K and $T = T_c$. The blue color stands for positive value and the red, negative.

## 5. Magneto-resistance contributed by superconducting vortex motion

For type-II superconductors, such as LSCO, superconducting vortices form when the superconductor is within a magnetic field, which is above the critical magnetic field $H_{c1}$. Each vortex is subject to the Lorentz force[35],

$$\boldsymbol{F}_L = \boldsymbol{\phi}_0 \times \boldsymbol{J},$$

where $\boldsymbol{J}$ is the current density and $\boldsymbol{\phi}_0$ is the flux quantum.

For our first measurement scheme, both $\boldsymbol{B}$ and $\boldsymbol{J}$ are in-plane, and the angle between them is $\varphi$. Thus, the Lorentz force is out-of-plane and proportional to $sin\varphi$. The induced vortex motion with the velocity, $\boldsymbol{v}$, is normal to the plane, and it cuts through the magnetic field, dissipating energy and generating an in-plane electric field,

$$\boldsymbol{E} = \boldsymbol{v} \times \boldsymbol{B}/c.$$

Therefore, the electric field longitudinal and transverse to $\boldsymbol{J}$, $E$ and $E_T$, have such dependencies on $\varphi$ as

$$E \propto sin\varphi \cdot \sin\varphi \propto -\cos 2\varphi,$$

$$E_T \propto sin\varphi \cdot \cos\varphi \propto \sin 2\varphi.$$

In accord, the longitudinal and transverse resistivities, $\rho(\phi = 0°, \varphi)$ and $\rho_T(\phi = 0°, \varphi)$, satisfy the relations:

$$\rho(\phi = 0°, \varphi) = \bar{\rho} - \Delta\rho \cos(2\varphi) \quad \ldots (3)$$

$$\rho_T(\phi = 0°, \varphi) = \bar{\rho}_T + \Delta\rho \sin(2\varphi) \quad \ldots (4)$$

Here $\bar{\rho}$ and $\bar{\rho}_T$ are the average longitudinal and transverse resistivities, respectively, and $\Delta\rho$ is a fitting parameter corresponding to the amplitude of angular oscillations. $\bar{\rho}_T$ is non-zero due to the presence of electronic nematicity at zero magnetic field.

Note that although Expressions (3) and (4) look similar to Expressions (1) and (2), the physics origins are drastically different: the former is due to the Lorentz force acted by the magnetic field; the latter is due to electronic nematicity of LSCO in the absence of the magnetic field.

The typical $\rho(\phi = 0°, \varphi)$ and $\rho_T(\phi = 0°, \varphi)$ in the superconducting fluctuating state can be well fitted by Expressions (3) and (4) (Fig. S6a). This confirms firmly that the observed angular oscillations using the first measurement scheme are contributed by superconducting vortex motions.

Furthermore, $\Delta\rho(B)$ at various temperatures are shown in Fig. S6b. For the superconducting state, e.g., $T$ = 38 K, both the population of superconducting vortex and the induced $E$ field are proportional to $B$. Consequently, $\Delta\rho$ increases with $B$. For the superconducting fluctuating state, e.g., $T$ = 40.5 K, the residual superconductivity is severely suppressed by $B$ so that $\Delta\rho$ decreases with $B$. The two competing effects work together in producing the non-monotonic $\Delta\rho(B)$ at 40 K.

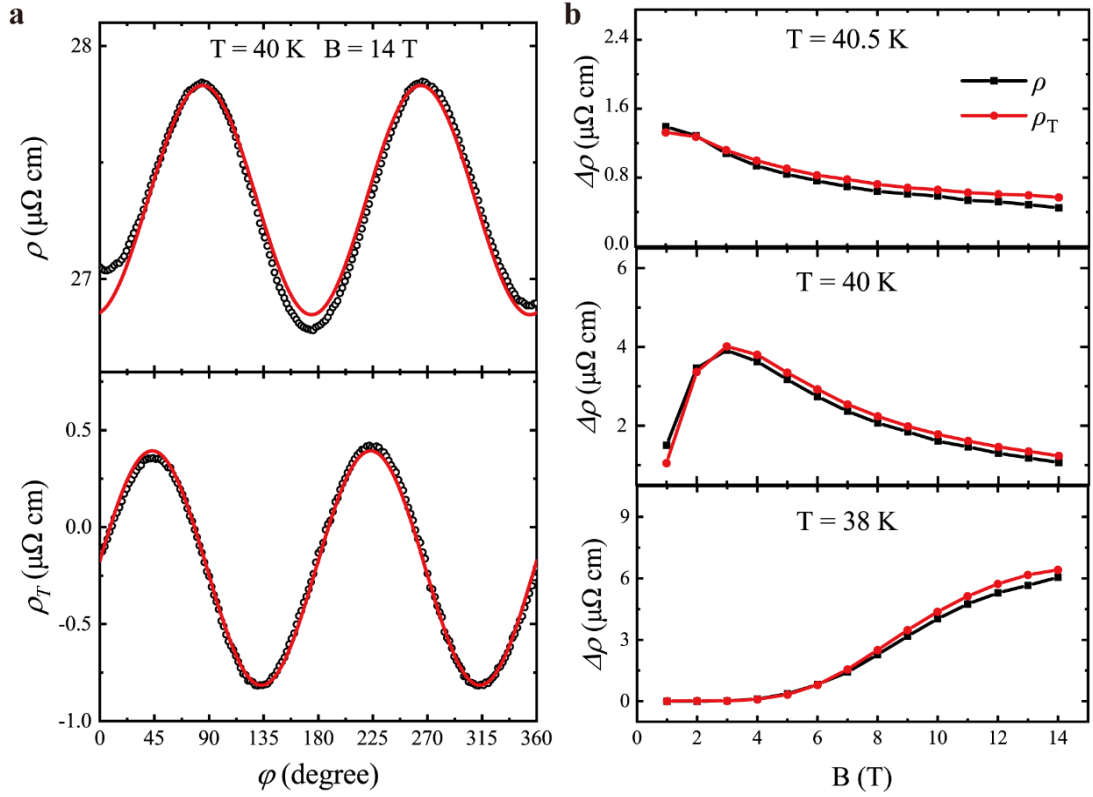

**Figure S6 | Dissipative superconducting vortex motion driven by the Lorentz force. a**, The longitudinal resistivity $\rho(\phi = 0°, \varphi)$ and transverse resistivity $\rho_T(\phi = 0°, \varphi)$ at $T$ = 40 K and $B$ = 14 T. The black open dots are experimental data and the red curves are the best fittings according to Expressions (3) and (4). **b**, The amplitude of angular oscillation, $\Delta\rho$, as a function of $B$ at various $T$. $\Delta\rho$ retrieved from $\rho(\phi = 0°, \varphi)$ (black curve) and $\rho_T(\phi = 0°, \varphi)$ (red curve)

## 6. Linear *I-V* relations above $T_c$

The *I-V* curves are non-linear if the excitation current is close to $I_c$, the critical current. This may interfere with the angle dependence of magneto-resistivity measurements. Here, we show representative *I-V* relations for both **B** // **J** and **B** ⊥ **J** measurement schemes at $T$ = 38.2 K and $B$ = 13 T, which corresponds to $\rho(\phi, \varphi = \phi)$ with **C₄** symmetry in the *B-T* phase diagram (Fig. 4). The current used for electrical transport measurement is $I = 100\mu A$, and it is apparent that the *I-V* curves are strictly linear from -400 $\mu A$ to 400 $\mu A$ (Fig. S7). Therefore, we can safely rule out the possibility that **C₄** symmetry in $\rho(\phi, \varphi = \phi)$ originates from non-linear *I-V* dependence.

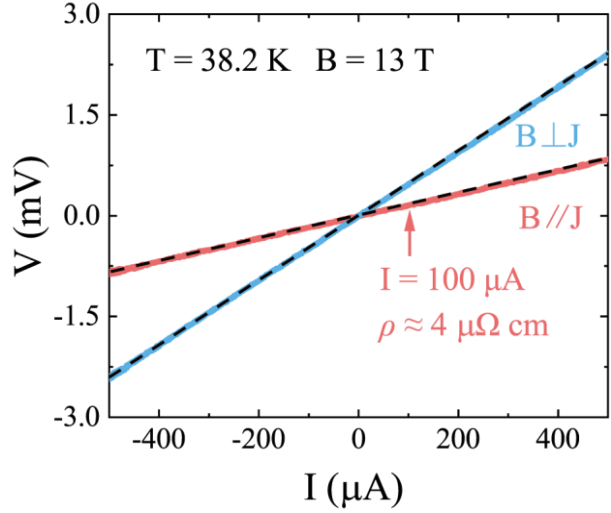

**Figure S7 │ I-V curves of LSCO film at T = 38.2 K and B = 13 T.** For both ***B*** // ***J*** and ***B*** ⊥ ***J*** measurement schemes (***J*** // LSCO [100]), the I-V curves are linear for $I = 100$ μA, which is the current used for electrical transport measurements. The solid curves are experimental data and the dashed lines are the linear fittings.

### 7. In-plane versus out-of-plane anisotropy

Superconductivity of LSCO is quasi-two-dimensional and the critical magnetic field is strongly anisotropic along the in-plane and out-of-plane directions. An example of $\rho(\theta)$ is shown in Fig. S8, where $\theta$ is the azimuth angle between ***B*** and the LSCO [100] direction. Apparently, at $T = 35.2$ K, the LSCO film is superconducting along the in-plane direction while it is resistive along the out-of-plane direction.

We took advantage of this strong anisotropy by using $\rho(\theta)$ as an indicator to calibrate the azimuth angle. By rotating the sample relative to the magnetic field while searching for the angle, at which $\rho(\theta)$ is the maximum, we can determine the angle $\theta = 0°$ very accurately. This guarantees the effect of the out-of-plane component of magnetic field is negligible in our measurements of in-plane magneto-resistivity.

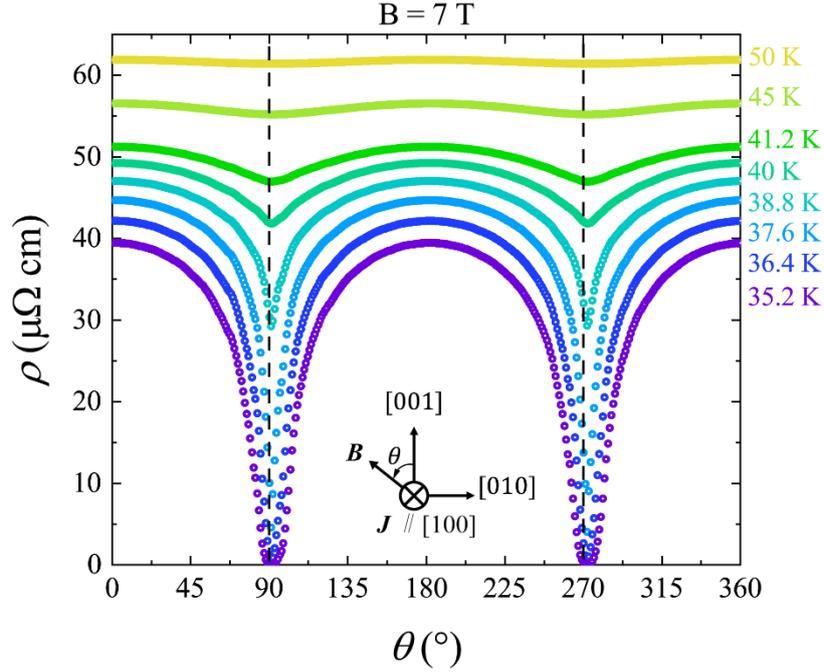

**Figure S8 | The magneto-resistance as a function of azimuth angle, $\theta$, between the magnetic field and the LSCO [001] direction.** The experiment was conducted on a LSCO($x = 0.16$) sample with $T_c^{off} = 38.3\ K$, which is slightly different from the one studied in the paper. The magnitude of the $B$ field remains 7 T and its direction rotates continuously from out-of-plane to in-plane.

## 8. Intrinsic C$_4$ symmetry in $\rho(\phi, \varphi = \phi)$

The **C$_4$** symmetry is intrinsic to LSCO films. Here, we show more examples of $\rho(\phi, \varphi = \phi)$ at fixed $B$ fields and varying temperatures (Fig. S9). A transition from fourfold to twofold angular oscillations occurs as $T$ increases, similar to $\rho(\phi, \varphi = \phi)$ with varying $B$ fields in Fig. 2d. This clearly demonstrates that both **C$_2$** and **C$_4$** symmetries are intrinsic to LSCO, supporting the conclusion that nematic and $d$-wave superconductive orders are generic and intertwined in copper-based high temperature superconductors.

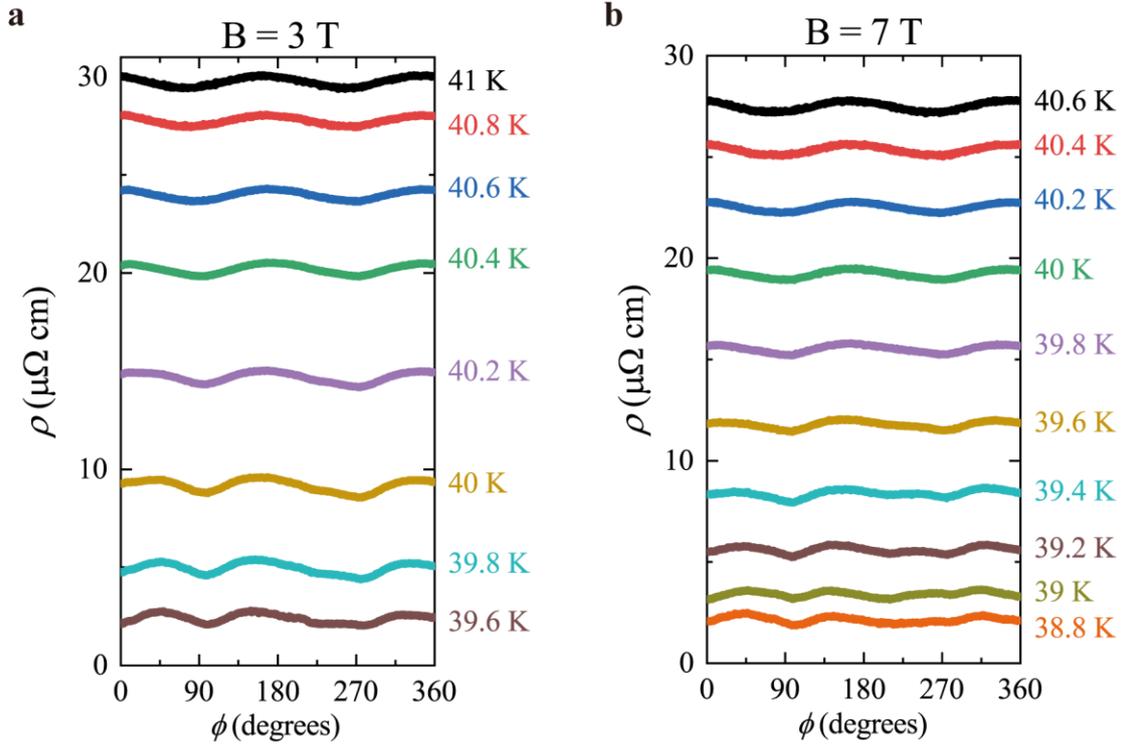

**Figure S9 | $\rho(\phi, \varphi = \phi)$ at fixed B fields and varying temperatures.** A transition from fourfold to twofold angular oscillations occurs as $T$ increases, similar to $\rho(\phi, \varphi = \phi)$ with varying $B$ fields in Fig. 2d. **a**, $B = 3$ T. **b**, $B = 7$ T.